\documentclass[11pt]{article}
\usepackage[utf8]{inputenc}
\usepackage[T1]{fontenc}
\usepackage{lmodern}
\usepackage[margin=1in]{geometry}
\usepackage{amsmath}
\usepackage{amssymb}
\usepackage{amsthm}
\usepackage{graphicx}
\usepackage{booktabs}
\usepackage{array}
\usepackage{siunitx}
\usepackage{xcolor}
\usepackage[colorlinks=true,linkcolor=black,citecolor=black,urlcolor=blue]{hyperref}
\usepackage[numbers,sort&compress]{natbib}
\usepackage{orcidlink}
\newcommand{\code}[1]{\texttt{#1}}
\graphicspath{{./}{figures/}}
\title{\textbf{Cislunar navigation is an observability problem: arc-length thresholds, Gramian conditioning, and the Doppler lever}}
\author{Chakshu Baweja\,\orcidlink{0009-0008-2098-0751}\\
\small Ashforde O\"U, Tallinn, Estonia \\
\small \texttt{contact@ashforde.org}}
\date{\today}
\begin{document}
\maketitle
\begin{abstract}
Dilution of precision is the standard way to describe navigation geometry and it
does not apply in cislunar space: its derivation presumes a snapshot solution in
which simultaneous measurements determine the state at one instant. A spacecraft tracked by inter-satellite ranging in the Earth-Moon three-body
problem has no such solution: range is a function of position only, so the
velocity columns of the measurement Jacobian vanish identically and the state
is unobservable however many links are in view. The correct figure of merit is dynamical, because the three-body flow carries
velocity information into later range measurements. We demonstrate this on four
differential-corrected planar distant retrograde orbits, reporting the
observability Gramian rather than a dilution of precision. Instantaneous range-only geometry
has rank \num{2} of \num{4}; the full state becomes observable only after a
tracking arc of \SI{2.09}{\hour}, the first sampled arc at full rank on a
\SI{0.26}{\hour} grid, and a square-root information filter recomputing the
same quantity by a different route becomes well-posed at exactly that arc.
Beyond it, conditioning improves monotonically with arc length, the Gramian
condition number falling from \num{1.2e11} at \SI{3}{\hour} to \num{1.4e5} at
\SI{48}{\hour}. Adding range rate lifts the \emph{instantaneous} rank from \num{2} to
\num{4}, so a Doppler-capable link makes a snapshot solution exist where a
range-only link cannot. Lifting the known-reference assumption, the joint problem of estimating a
station and the satellites' states and clocks together is rank-deficient by
exactly one without an external interferometric observable in all twenty
realizations tested, and full rank with it; the deficiency persists only to
four satellites. For the configurations studied, cislunar navigation requirements are better
written as arc-length, conditioning and measurement-type budgets than as
dilution-of-precision thresholds, which are undefined here.
\end{abstract}
\section{Introduction}
\label{sec:intro}

Ask a navigation engineer how good a geometry is and the answer will be a
dilution-of-precision number. The concept is so embedded that it is often
treated as a property of a constellation rather than of a solution method. It
is in fact the latter, and the distinction becomes decisive in cislunar space.

Dilution of precision is defined for a snapshot solution. One assumes that at
a single epoch the measurements available determine the unknown state, linearizes
about a nominal, and reports a scalar summary of the resulting covariance scaled
by the measurement sigma. The assumption that the instant is sufficient is doing
the work. For a terrestrial user solving for three position components and a
clock offset from four or more simultaneous pseudoranges, it holds. For a
spacecraft in cislunar space being tracked by inter-satellite ranging, it fails,
and it fails structurally rather than marginally.

The reason is elementary once stated. A range measurement between two
spacecraft depends on their positions and not on their velocities. If the
unknown state includes velocity, as it must for an orbiting body, then every
row of the instantaneous measurement Jacobian has identically zero entries in
the velocity columns. The instantaneous Jacobian therefore cannot have full
rank with respect to the full state, regardless of how many spacecraft are in
view or how favorably they are placed. Adding links improves nothing about this;
the deficiency is in the measurement model, not the geometry. A dilution of
precision computed for such a configuration is not a large number. It is
undefined, and a pipeline that reports one is inverting a singular matrix with
the singularity hidden by numerical tolerance.

\subsection{What replaces it}

Velocity is recoverable, but only through time. The dynamics propagate the
state, so a range measurement taken later is a function of the earlier
velocity. The information about velocity is therefore carried by the flow, and
the right question is not how the spacecraft are arranged at an instant but how
much information the trajectory accumulates over an arc. That question is
answered by the observability Gramian, which for a linearized system is
assembled from the measurement Jacobians mapped back to an epoch through the
state transition matrix. Its rank says whether the state is recoverable at all;
its condition number says how well.

This substitution is not merely notational. It changes what a requirement looks
like. A dilution-of-precision threshold is a statement about instantaneous
geometry that a mission can check continuously. An observability requirement is
a statement about \emph{arc length} and \emph{conditioning}, which constrains
tracking schedules, contact plans and maneuver cadence instead. Getting this
right matters for the lunar navigation services now being designed, because the
relays that will broadcast to surface users must themselves be determined, and
the accuracy with which they can be determined sets the ranging accuracy those
users receive \citep{baweja2026integrity}.

\subsection{Contribution}

\begin{enumerate}
\item \textbf{A quantified arc-length threshold on real orbits.} On a
four-spacecraft constellation of differential-corrected planar distant
retrograde orbits, the observable subspace has rank \num{2} of \num{4}
instantaneously and reaches full rank only at an arc of \SI{2.09}{\hour}
(Section~\ref{sec:results}). The orbits are genuine periodic solutions, closed
to periodicity residuals between \num{3.8e-11} and \num{4.7e-9}, not bounded
arcs that merely look like orbits.

\item \textbf{The joint problem, where the references are unknown too.}
Lifting the known-reference assumption (Section~\ref{sec:joint}) sharpens the
result rather than softening it. Estimating a surface station and the
constellation members' states and clocks together is rank-deficient by exactly
one without an external interferometric observable, in every one of twenty
realizations, and full rank with it. The deficiency is confined to small
constellations, persisting to four satellites and vanishing from five, which
makes an Earth-baseline interferometric capability structurally necessary
precisely in the near-term Artemis regime and optional beyond it.

\item \textbf{A second numerical route to the same threshold.} A square-root
information filter, processing the same measurements sequentially rather than in
batch, becomes well-posed at exactly the same arc, with condition numbers
matching the Gramian to six significant figures. Because both routes consume the
same Jacobians, this is a consistency check rather than independent
corroboration; it rules out an implementation artifact in either route, and we
claim no more than that.

\item \textbf{Conditioning as the design quantity, with its scale.} Beyond the
rank threshold, the Gramian condition number falls monotonically with arc
length, from \num{1.2e11} immediately past the threshold at \SI{3}{\hour} to
\num{1.4e5} at \SI{48}{\hour}. Rank tells a mission whether a solution exists; conditioning
tells it whether the solution is worth having, and only the second is a useful
design axis.

\item \textbf{The Doppler lever.} Augmenting the same links with range-rate
lifts the instantaneous rank from \num{2} to the full \num{4}. A Doppler
measurement is a function of velocity, so it populates the columns that range
alone leaves empty. This converts a problem that requires a multi-hour arc into
one admitting a snapshot solution, and it is a cheaper lever than adding
spacecraft.
\end{enumerate}

\subsection{Relation to the static case}

The observability lens applied here has a static counterpart. For fixed
surface stations the analogous question is whether the geodetic datum itself is
determined, and the answer is governed by a rank-deficiency of the estimation
problem whose null space corresponds to unresolvable frame transformations
\citep{baweja2026lunarpnt}. The present paper carries that lens from the static
configuration into the dynamical one. The two treat different unknowns, fixed
station coordinates against a moving spacecraft's full state, and different
mechanisms for the deficiency, a frame null space against absent velocity
columns. This is a generalization of the viewpoint rather than a re-derivation
of the result.

\subsection{Validated versus modelled}

\textbf{Validated.} The three-body dynamics, the variational state transition
matrix and the differential corrector are those of the host engine, whose state
transition matrix is verified against finite differences and whose corrected
periodic orbits conserve the Jacobi constant; the periodicity residuals are
reported per orbit in Section~\ref{sec:results}. The rank and conditioning
results are cross-checked against a second, sequential estimator.
\textbf{Modelled.} The dynamics are the planar circular restricted three-body
problem with a single mass parameter, omitting solar perturbation, eccentricity
of the lunar orbit, and non-spherical gravity. Measurements are noise-free
geometric ranges and range rates with no light-time, clock, or bias modelling.
The consequence is that the rank structure and the arc thresholds are properties
of the dynamics and the measurement model and are robust, while the specific
condition-number magnitudes are indicative of the regime rather than predictions
for a particular mission.
\section{Related work}
\label{sec:related}

\paragraph{Autonomous navigation from inter-satellite range.} That two
spacecraft can be jointly determined from the scalar range between them, with
no external reference, is the linked autonomous navigation concept
\citep{hill2008liaison}. It works because asymmetry in the gravitational field
breaks the scale and orientation degeneracies that make the same problem
unsolvable in the two-body case, which is precisely why cislunar space is a
favorable regime for it. The concept has been developed for lunar far-side
users and for distant retrograde orbits specifically
\citep{hesar2015liaison,wang2019liaison}.

\paragraph{Observability analysis in the orbital context.} The dynamical
perspective on when inter-satellite range suffices, and what in the dynamics
supplies the missing information, has been analysed directly
\citep{li2020observability}. That work establishes the mechanism we rely on:
observability in this problem is a property of the dynamics acting over time,
not of the instantaneous configuration. Our contribution relative to it is not
the mechanism but the operational framing, namely reporting an arc-length
threshold and a conditioning curve as the design quantities that should replace
a dilution-of-precision threshold, validating the threshold against a
second, sequential estimator, and identifying the range-rate lever that removes the
arc requirement altogether.

\paragraph{Navigation performance of the same orbit families.}
Achieved navigation performance for distant retrograde and other cislunar
orbits has been characterized, including jointly via linked autonomous
navigation with dynamic and clock model errors \citep{wang2019dro}, and lunar
integrity augmentation has been designed against the poor estimator
observability these geometries produce \citep{engwerda2026augment}. Those
studies report how well an estimator performs on such an orbit. The question
here is prior to performance: whether the state is recoverable at all from a
given measurement set, and over what arc. A performance figure presumes
observability, and this paper locates the threshold at which that presumption
becomes valid.

\paragraph{Cislunar constellation design.} Studies of cislunar navigation
constellations and their orbit determination performance are an active area,
and lunar navigation architectures are being specified now
\citep{iiyama2023lupnt,israel2020lunanet}. Where dilution of precision is
reported for such architectures, it is typically computed for a surface or
orbital user receiving from a constellation whose own states are assumed known.
That is a legitimate snapshot problem. The distinct problem treated here is the
determination of the constellation members themselves from links among them,
where the snapshot formulation is unavailable.

\paragraph{Estimation and conditioning.} The square-root information filter used as a second computational route is standard, and the numerical-conditioning
interpretation of the information matrix and its spectrum is classical
\citep{golub2013matrix,kay1993}. We use conditioning in its standard sense: the
ratio of extreme eigenvalues of the information Gramian bounds the amplification
of measurement error into state error along the worst-determined direction.

\paragraph{Companion analyses.} The integrity requirement that lunar relays
must ultimately satisfy is derived in \citep{baweja2026integrity}, and the
differential architecture that relaxes it is treated in
\citep{baweja2026dpnt}. The static-case datum observability argument this paper
generalizes is developed in \citep{baweja2026lunarpnt}. Broader program and
toolchain context is in \citep{baweja2026kshana,montenbruck2000,vallado2006}.

\paragraph{Autonomous determination from inter-satellite tracking.} The idea
that two or more spacecraft can be determined jointly from the ranges between
them, without an external frame, originates in satellite-to-satellite tracking
work \citep{hillborn2007} and matured into the linked autonomous navigation
concept \citep{hill2008liaison,hesar2015liaison,wang2019liaison}. That
literature establishes the mechanism we rely on, namely that asymmetry in the
three-body dynamics breaks the scale and orientation degeneracy a two-body
range-only network would suffer. \citet{li2020observability} analyze
observability of inter-satellite range navigation from an orbital-dynamics
perspective, which is the closest prior work to Section~\ref{sec:results}.
Related constellation-level work uses inter-satellite links for fault and clock
monitoring rather than state determination, using rigidity arguments
\citep{iiyama2024} and ranging-based clock fault detection
\citep{iiyama2025edm}, and lunar orbit and clock estimation has been treated
jointly by \citet{iiyama2026odclock}.

\paragraph{Observability, rigidity and design of experiments.} The question of
whether a network of range measurements determines its own configuration is the
subject of network localization theory \citep{aspnes2006}, which supplies the
rigidity language that the rank arguments here specialize to a dynamical
setting. The information matrix and its spectrum, the objects we report in place
of a dilution of precision, are standard in estimation theory
\citep{kay1993,teunissen2000} and in matrix analysis
\citep{hornjohnson2013matrix,golub2013matrix,zhang2005schur}. Reporting a
scalar summary of an information matrix, such as the smallest eigenvalue we use
as an E-optimality proxy, is the subject of optimal design of experiments
\citep{pukelsheim2006oed}, and the underlying convex-analytic framing is
standard \citep{boyd2004convex}. What this paper adds is the observation that in
the cislunar case the quantity conventionally reported, dilution of precision,
is not merely a poor summary but undefined, because the matrix it inverts is
singular.

\paragraph{External observables and tooling.} The interferometric observable
whose structural role Section~\ref{sec:joint} isolates is the deep-space
analogue of delta-differential one-way ranging, whose nanoradian-class
performance is described by \citet{curkendall2013deltador}. Reference
ephemerides for the Earth-Moon system are standard \citep{park2021de440}, and
open astrodynamics toolkits exist that implement the propagation and
differential-correction machinery used here
\citep{nyxspace2025,orekit,baweja2026kshana}.

\section{Method}
\label{sec:method}

\subsection{Dynamics}

We work in the planar circular restricted three-body problem in the rotating
Earth-Moon frame, nondimensionalized so that the primary separation, the total
mass and the mean motion are unity. With mass parameter
\(\mu = \num{0.01215}\), the state \(\mathbf{x} = [x, y, \dot{x}, \dot{y}]^{\top}\)
evolves as
\begin{equation}
\ddot{x} - 2\dot{y} = \frac{\partial \Omega}{\partial x}, \qquad
\ddot{y} + 2\dot{x} = \frac{\partial \Omega}{\partial y}, \qquad
\Omega = \tfrac{1}{2}(x^2+y^2) + \frac{1-\mu}{r_1} + \frac{\mu}{r_2},
\label{eq:cr3bp}
\end{equation}
with \(r_1, r_2\) the distances to the Earth and Moon. Linearized
perturbations of Equation~\eqref{eq:cr3bp} propagate through the state
transition matrix \(\Phi(t, t_0)\)
obtained by integrating the variational equations
\(\dot{\Phi} = A(t)\,\Phi\), \(\Phi(t_0,t_0) = I\), alongside the trajectory.

\subsection{Constellation}

The four spacecraft occupy planar distant retrograde orbits about the Moon,
each obtained by differential correction: a state on the \(x\) axis with purely
transverse velocity is propagated to the next perpendicular axis crossing and
the velocity adjusted until the crossing is perpendicular, which by symmetry
yields a closed periodic orbit. We report the resulting periodicity residual,
the distance between the initial and final state after one full period, for
each member as evidence that the orbits are genuinely periodic rather than
merely bounded. One member is designated the chief, whose state is to be
determined; the other three are references at known states, phased around their
orbits.

\subsection{Measurements and the rank deficiency}

The range from the chief at \(\mathbf{r}\) to reference \(i\) at
\(\mathbf{p}_i\) is \(\rho_i = \lVert \mathbf{r} - \mathbf{p}_i \rVert\), whose
Jacobian with respect to the full state is
\begin{equation}
\frac{\partial \rho_i}{\partial \mathbf{x}}
= \left[\; \hat{\mathbf{u}}_i^{\top} \;\middle|\; \mathbf{0}^{\top} \;\right],
\qquad
\hat{\mathbf{u}}_i = \frac{\mathbf{r}-\mathbf{p}_i}{\lVert \mathbf{r}-\mathbf{p}_i\rVert},
\label{eq:rangejac}
\end{equation}
where the zero block occupies the velocity columns. Stacking any number of such
rows at one epoch gives a matrix of rank at most \num{2} in the planar case,
never \num{4}: the deficiency is structural. The range rate
\(\dot{\rho}_i = \hat{\mathbf{u}}_i^{\top}(\dot{\mathbf{r}} - \dot{\mathbf{p}}_i)\)
has a Jacobian whose velocity block is \(\hat{\mathbf{u}}_i^{\top}\) and is
therefore not zero, which is the mechanism behind the Doppler lever of
Section~\ref{sec:results}.

\subsection{Observability Gramian}

Mapping each epoch's measurement Jacobian \(H_k\) back to the initial epoch
through the state transition matrix and accumulating gives the discrete
observability Gramian
\begin{equation}
\mathcal{G}(t_0, t_f) \;=\; \sum_{k} \Phi(t_k,t_0)^{\top} H_k^{\top} H_k\, \Phi(t_k,t_0),
\label{eq:gramian}
\end{equation}
whose rank is the dimension of the observable subspace and whose condition
number \(\kappa(\mathcal{G}) = \lambda_{\max}/\lambda_{\min}\) measures how
unevenly information is distributed across state directions. Rank is evaluated
by singular-value decomposition with a relative tolerance; we report the
smallest singular value alongside the rank so that the threshold is not an
artifact of the tolerance.

\subsection{Cross-check by a second estimator}

Because a rank threshold computed by one numerical route can be an artifact of
that route, we recompute the same quantity by a different one. A square-root
information filter processes the same measurements sequentially, propagating an
upper-triangular information square root; its posterior is well-posed exactly
when the accumulated information matrix is invertible. The filter shares no
code with the Gramian assembly of Equation~\eqref{eq:gramian}. Agreement
between the arc at which the Gramian attains full rank and the arc at which the
filter posterior becomes well-posed is therefore meaningful evidence, and
disagreement would have been a red flag.

\section{Results}
\label{sec:results}

\subsection{The orbits are genuinely periodic}

Because the entire argument rests on the dynamics rather than on geometry, the
trajectories must actually be the orbits they are claimed to be. Each member
was differentially corrected to a perpendicular axis crossing and then
propagated through a full period; Table~\ref{tab:dro} reports the resulting
periodicity residual, the separation between the initial and final state. The
residuals lie between \num{3.8e-11} and \num{4.7e-9} in nondimensional units,
which is closure at the level of the integrator rather than approximate
boundedness. The four orbits span perilune radii from
\SI{18345}{\kilo\meter} to \SI{44120}{\kilo\meter} and periods from
\num{0.55} to \num{1.82} nondimensional time units, so the constellation is a
genuinely nested family rather than four near-copies.

\begin{table}[t]
\centering
\caption{The four differential-corrected planar distant retrograde orbits. The
periodicity residual is the nondimensional distance between the initial and
final state after one full period, and is the evidence that these are closed
periodic solutions rather than bounded arcs.}
\label{tab:dro}
\begin{tabular}{l S[table-format=5.0] S[table-format=1.3] S[table-format=1.2e-2] S[table-format=1.2]}
\toprule
{role} & {perilune (km)} & {period (TU)} & {periodicity residual} & {phase} \\
\midrule
chief       & 26589 & 0.923 & 6.06e-10 & 0.00 \\
reference 1 & 18345 & 0.551 & 4.73e-09 & 0.20 \\
reference 2 & 35794 & 1.380 & 1.21e-10 & 0.44 \\
reference 3 & 44120 & 1.818 & 3.85e-11 & 0.66 \\
\bottomrule
\end{tabular}
\end{table}

\subsection{A snapshot cannot solve the problem, and an arc can}

The instantaneous measurement Jacobian built from all three range links has
rank \num{2} against a four-dimensional state, exactly as
Equation~\eqref{eq:rangejac} predicts: the two velocity columns are identically
zero and cannot be populated by any number of simultaneous range measurements.
A dilution of precision for this configuration does not exist. This is worth
stating carefully, because it is not the claim that the geometry is poor. It is
the claim that the quantity conventionally used to describe geometry is
undefined here, and that a pipeline reporting a large but finite value for it is
reporting the reciprocal of a numerical tolerance.

Accumulating measurements along the trajectory changes this.
Figure~\ref{fig:main}(a) shows the observable rank against arc length, sampled
on a fine per-epoch grid of step \SI{0.26}{\hour}. The rank climbs as the flow
carries velocity information into successive range measurements. The last
sampled arc at which the Gramian is rank-deficient is \SI{1.83}{\hour} and the
first at which it attains full rank is \SI{2.09}{\hour}, so the transition lies
in the interval \(( \SI{1.83}{\hour},\, \SI{2.09}{\hour} ]\). We quote
\SI{2.09}{\hour} throughout as the first sampled full-rank arc, not as a
root-found threshold, and the two-figure precision reflects the grid rather than
a resolved crossing. The smallest singular value at that arc is \num{3.87e-6}
and grows steadily thereafter, reaching \num{1.47e-5} by \SI{3.13}{\hour}, so
the transition is not an artifact of the rank tolerance.

\subsection{A second numerical route agrees, exactly}

A rank threshold obtained from one numerical procedure invites the suspicion
that it is an artifact of the procedure. We therefore recomputed it by a
different route: a square-root information filter that processes the same
measurements sequentially and maintains an information square root, rather than
assembling and decomposing the Gramian in batch. We are explicit about what this
does and does not establish. The filter consumes the same measurement Jacobians
and state transition matrices, so the information content is by construction the
same object; agreement is therefore a consistency check between two computations
of one quantity, not independent physical corroboration, and the
six-significant-figure match is close to a numerical identity. What it does rule
out is an implementation or conditioning artifact in either route, which is a
real and worthwhile check but a narrower claim than independence. The
filter posterior becomes well-posed at \SI{2.09}{\hour}, the same arc, and its
condition number at that epoch is \num{6.025575e11} against the Gramian's
\num{6.025578e11}, agreeing to six significant figures; the agreement persists
across the arc, for example \num{3.094469e11} against \num{3.094469e11} at the
next epoch. Genuine independent corroboration would require an estimator with a
different measurement parameterization, or flight data; neither is available
here, and we therefore treat \SI{2.09}{\hour} as a robust numerical result for
this configuration rather than as a validated property of the physical problem.

\subsection{Conditioning is the design axis, and arc length buys it}

\begin{figure}[t]
\centering
\includegraphics[width=\textwidth]{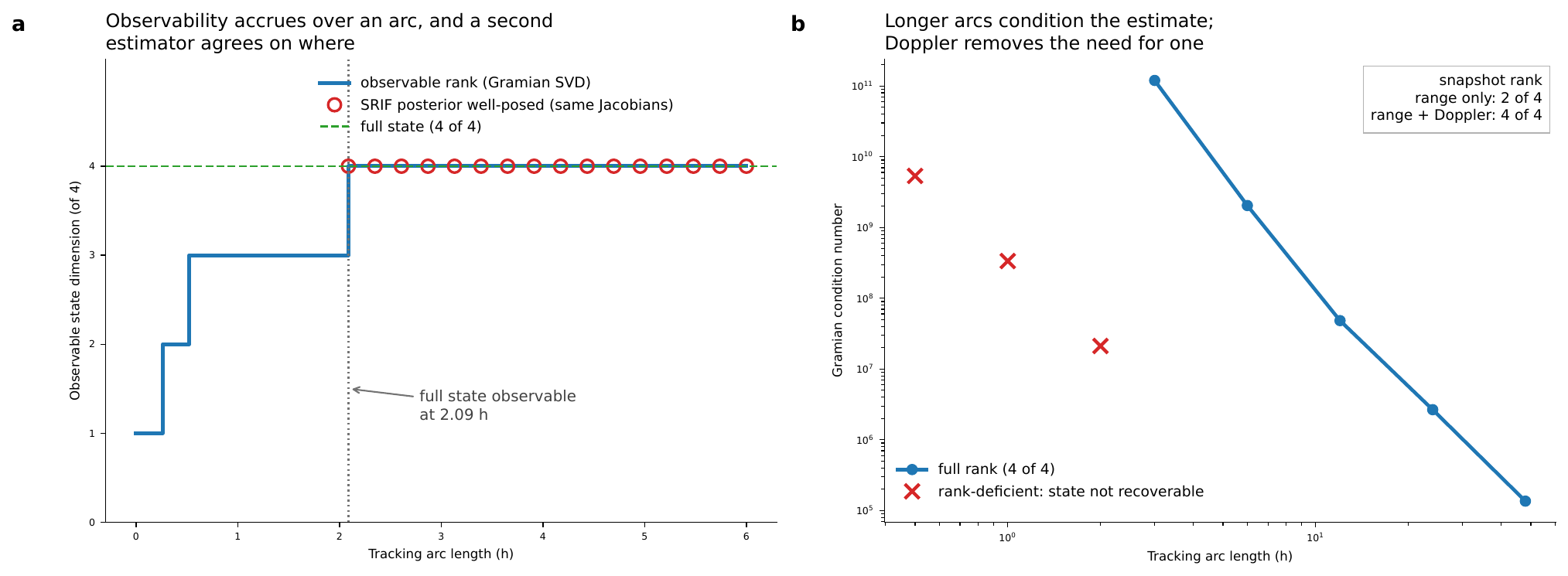}
\caption{\textbf{Cislunar observability is dynamical, not geometric.}
(a) Observable state dimension against tracking arc length for the
four-spacecraft distant-retrograde constellation. Range-only measurements leave
the state unobservable at any instant; rank accumulates through the three-body
flow and reaches the full dimension \num{4} at \SI{2.09}{\hour}. Open circles
mark the arcs at which a square-root information filter posterior becomes well-posed, which is the same arc. (b) Gramian condition number against
arc length on logarithmic axes, for the full-rank arcs (filled); crosses mark
rank-deficient arcs, whose condition number is taken over the observable
subspace only and is not a usable figure of merit. Conditioning improves by
about six orders of magnitude between \SI{3}{\hour} and \SI{48}{\hour}. The
inset states the lever: adding range rate to the same links lifts the
instantaneous rank from \num{2} to \num{4}, making a snapshot solution exist.}
\label{fig:main}
\end{figure}

Rank is binary and therefore a poor requirement. Table~\ref{tab:arc} and
Figure~\ref{fig:main}(b) report the conditioning. Immediately past the
threshold the Gramian is extremely ill-conditioned, with
\(\kappa = \num{1.2e11}\) at \SI{3}{\hour}: the fourth direction is technically
observable but barely so, and an estimator operating there would amplify
measurement error enormously along it. Conditioning then improves monotonically,
to \num{2.0e9} at \SI{6}{\hour}, \num{4.8e7} at \SI{12}{\hour},
\num{2.7e6} at \SI{24}{\hour} and \num{1.4e5} at \SI{48}{\hour}. Extending the
tracking arc from \SI{3}{\hour} to \SI{48}{\hour} buys roughly six orders of
magnitude in conditioning without adding a single spacecraft or improving a
single sensor.

This is why an operational requirement should be written on conditioning rather
than rank. A mission that satisfies "the state is observable" at
\SI{2.1}{\hour} has satisfied nothing useful; a mission that specifies a
conditioning target has specified a tracking schedule.

\begin{table}[t]
\centering
\caption{Observability against tracking arc length. Below \SI{2.09}{\hour} the
Gramian is rank-deficient and the state is not recoverable; the condition
number quoted for those rows is taken over the observable subspace only and is
reported for continuity rather than as a usable figure of merit. The
instantaneous range-only and range-plus-Doppler ranks are reported separately in
Section~\ref{sec:results} and are not tabulated here.}
\label{tab:arc}
\begin{tabular}{S[table-format=2.1] c c S[table-format=1.2e-2]}
\toprule
{arc (h)} & {Gramian rank} & {defect} & {condition number} \\
\midrule
0.5  & 3 & 1 & 5.36e9 \\
1.0  & 3 & 1 & 3.36e8 \\
2.0  & 3 & 1 & 2.12e7 \\
3.0  & 4 & 0 & 1.20e11 \\
6.0  & 4 & 0 & 2.05e9 \\
12.0 & 4 & 0 & 4.83e7 \\
24.0 & 4 & 0 & 2.66e6 \\
48.0 & 4 & 0 & 1.36e5 \\
\bottomrule
\end{tabular}
\end{table}

\subsection{The Doppler lever removes the arc requirement}

Everything above concerns range-only tracking. Augmenting the same three links
with range rate changes the instantaneous problem qualitatively. Because the
range-rate Jacobian has a nonzero velocity block, three simultaneous
range-and-range-rate measurements give an instantaneous Jacobian of rank
\num{4}: the full planar state is recoverable from a single epoch. We verified
this at every arc in the sweep, where the range-only instantaneous rank is
\num{2} and the range-plus-Doppler instantaneous rank is \num{4} without
exception.

The engineering consequence is direct and, we think, underappreciated. The
choice between a range-only and a Doppler-capable inter-satellite link is not a
choice between two accuracies. It is the choice between a navigation problem
that requires hours of continuous tracking before a solution exists and one
that admits a snapshot solution. Against that, adding more range-only
spacecraft buys nothing structural: it improves the conditioning of a problem
that remains rank-deficient at any instant.

\section{The joint problem: when the references are unknown too}
\label{sec:joint}

Section~\ref{sec:results} treats the reference spacecraft as having known
states. That is the standard first step, and it is also the most serious
limitation of the analysis, because the operationally relevant cislunar problem
is the joint one in which nothing is known: the constellation members, their
clocks, and any surface or orbital user are estimated together from the same
measurements. This section addresses that case, and finds that the
observability lens gives a sharper answer there than it does in the
known-reference case.

\subsection{Setup}

We estimate a joint state comprising a lunar surface station's position and
clock offset together with the positions and clock offsets of \(n_{\mathrm{sat}}\)
lunar satellites, from inter-satellite ranges, Earth-based ranges from
\(n_{\oplus}\) ground stations, and optionally a very-long-baseline
interferometric delay observable. The default configuration has
\(n_{\mathrm{sat}} = 3\) and \(n_{\oplus} = 6\), giving \num{16} estimated
parameters. We report the rank and defect of the information matrix, the
Cramer-Rao lower bound on station position where it exists, the fraction of
station position axes that are unobservable, and the realized errors of a
converged least-squares solution. Every configuration is run both with and
without the interferometric observable, so the comparison isolates what that
one measurement type contributes.

\subsection{Without an external link, the joint problem is rank-deficient}

The central result is structural. Without the interferometric observable, the
joint information matrix has rank \num{15} of \num{16}: a defect of exactly
one. The deficiency is not a conditioning complaint but a genuine null
direction, and it is concentrated in station position. We
quantify that concentration as the squared norm of the projection of the
information matrix's null vector onto the three station-position coordinates,
which is \num{0.34}: roughly a third of the single unobservable direction lies
in station position and the remainder is spread across the satellite states and
clocks. The measure is a fraction of one direction, not a count of axes. Adding the interferometric observable
restores full rank \num{16} and drives the unobservable-axis measure to exactly
zero.

We checked that this is a property of the problem rather than of one noise
draw. Across twenty independent realizations the defect without the external
link is exactly one in every case, and exactly zero with it in every case
(Table~\ref{tab:joint}). The rank structure never varies. This is the same
observational logic as Section~\ref{sec:results}: a measurement set can be
arbitrarily precise and still leave a direction of the state unrecoverable, and
no amount of averaging fixes it.

\subsection{The magnitude of the benefit is highly realization-dependent}

The realized improvement is a different matter, and it is where a single run
would badly mislead. Across the same twenty realizations the station-position
improvement factor from adding the interferometric observable has a median of
\num{83} but ranges from \num{0.9} to \num{2395}. A single draw of this
experiment could report a three-order-of-magnitude benefit or no benefit at
all, and both would be true of that draw.

The reason is visible in the structure. When the information matrix is
rank-deficient, the least-squares solution is determined by the regularization
and the particular noise realization rather than by the data, so its error is
essentially arbitrary: the without-link station error ranges over
\SIrange{58}{3619}{\meter} across the twenty draws. The with-link error is far
better behaved, at \SIrange{0.6}{63}{\meter}. The honest statement is therefore
that the external link changes the problem from ill-posed to well-posed, which
is categorical, and that the resulting accuracy gain is typically around two
orders of magnitude but is not a reliable single number.

\begin{table}[t]
\centering
\caption{Joint estimation of station and satellite states and clocks, twenty
independent realizations at \(n_{\mathrm{sat}} = 3\), \(n_{\oplus} = 6\),
\num{16} parameters. The rank structure is invariant across realizations; the
realized errors are not.}
\label{tab:joint}
\begin{tabular}{l c c}
\toprule
& {without external link} & {with external link} \\
\midrule
information-matrix rank & 15 of 16 & 16 of 16 \\
defect (all 20 draws) & 1 & 0 \\
unobservable station axes & 0.34 & 0.00 \\
station error, median & \SI{900}{\meter} & \SI{13.3}{\meter} \\
station error, range & \SIrange{58}{3619}{\meter} & \SIrange{0.6}{63}{\meter} \\
improvement, median & \multicolumn{2}{c}{\num{83}} \\
improvement, range & \multicolumn{2}{c}{\numrange{0.9}{2395}} \\
\bottomrule
\end{tabular}
\end{table}

\subsection{The deficiency exists only for small constellations}

Sweeping the number of satellites locates the boundary precisely, and the
answer qualifies the result in a way that matters for mission design. The
defect without the external link is \num{2} at two satellites, \num{1} at three
and at four, and \num{0} from five satellites upward. Beyond four satellites
the inter-satellite geometry alone makes the joint problem observable, and the
external link stops being structurally necessary. Consistent with that, its
measured benefit collapses at the same point: from \num{615} at three
satellites and \num{59} at four to \num{0.76} at five, where adding an extra
noisy observable to an already well-posed problem slightly degrades the
particular solution rather than improving it.

The design implication is specific rather than general. An external
interferometric link is not universally valuable; it is structurally necessary
exactly in the regime of small cislunar constellations, which is the near-term
Artemis-era regime, and becomes optional once the constellation exceeds four
members. A programme deciding whether to fund an Earth-baseline interferometric
capability should therefore tie that decision to the planned constellation
size, and the transition is sharp rather than gradual.

We also note a floor on the ground segment: at two Earth stations the joint
solution is unobservable with or without the external link, and three are
needed before any configuration converges.

\subsection{Link precision sets the magnitude, not the rank}

Finally, the precision of the external observable trades against its value in
the expected way while leaving the structural result untouched. Sweeping the
interferometric noise from \SI{1}{\pico\second} to \SI{10}{\nano\second}, the
median improvement falls from \num{4665} to \num{17}, and the station
Cramer-Rao bound rises from \SI{2.5}{\meter} to \SI{91}{\meter}. The defect
without the link remains exactly one at every precision. Rank is determined by
which measurement types are present, and precision only determines how much
the resulting estimate is worth; this is the same separation between
existence and quality that Section~\ref{sec:results} draws between rank and
conditioning.

\section{Discussion}
\label{sec:discussion}

\subsection{What a cislunar navigation requirement should say}

The practical output of this analysis is a change in the form of the
requirement, not merely in its value. A dilution-of-precision threshold is
unusable here because the quantity does not exist for range-only tracking of a
moving body. Three replacements follow from Section~\ref{sec:results}.

First, an \emph{arc-length} requirement, stating the minimum continuous
tracking span before the state is recoverable at all. For the constellation
studied that is \SI{2.09}{\hour}, and it constrains contact scheduling and
maneuver cadence: a maneuver resets the arc, so maneuvers spaced more closely
than the threshold leave the state permanently unobservable.

Second, a \emph{conditioning} requirement, because rank alone is nearly
worthless operationally. The difference between an arc of \SI{3}{\hour} and one
of \SI{48}{\hour} is six orders of magnitude in the amplification of
measurement error along the worst-determined direction, and both satisfy a rank
requirement identically.

Third, a \emph{measurement-type} requirement, which the Doppler result shows is
the highest-leverage of the three. Specifying range-and-range-rate rather than
range-only changes the instantaneous rank from \num{2} to \num{4} and thereby
removes the arc-length requirement entirely.

\subsection{Why this matters to the surface user}

The chain runs from this paper to the surface. A lunar navigation service
broadcasts ephemeris and clock to users, and the accuracy of that broadcast is
limited by how well the relays' own states are determined, which is the problem
analysed here. The companion integrity analysis shows that a surface user needs
a signal-in-space ranging accuracy of a few meters for a trustworthy protection
level \citep{baweja2026integrity}, and that requirement cannot be met if the
relays are themselves determined through an ill-conditioned estimate. The
conditioning curve of Figure~\ref{fig:main}(b) is therefore not an abstract
numerical property; it is an upstream constraint on the broadcast error budget
that the surface user ultimately inherits.

\subsection{One lens, three different deficiencies}

Taken together the results describe the same diagnostic applied at three
levels, and it is worth naming them side by side because each has a different
remedy.

At the level of a single epoch with known references, the deficiency is in the
velocity subspace and the remedy is \emph{time}: the three-body flow carries
velocity information into later range measurements, and full rank arrives after
a tracking arc. At the level of a single epoch with a richer measurement set,
the deficiency is in the measurement model and the remedy is \emph{observable
type}: range rate makes the velocity columns non-zero, so a snapshot solution
exists where a range-only snapshot cannot. At the level of the joint problem
with unknown references, the deficiency is in station position and the remedy
is an \emph{external link} that ties an internally self-consistent cislunar
network to an outside frame.

None of the three is a geometry problem in the dilution-of-precision sense, and
no dilution-of-precision number would have diagnosed any of them, because in
each case the quantity DOP is computed from does not exist. That is the
argument of this paper in one paragraph: the first question about a cislunar
navigation configuration is not how good the geometry is but whether the
estimation problem is well posed, and the answer decomposes into time,
measurement type and external reference.

\subsection{Limitations}

The dynamics are the planar circular restricted three-body problem. The planar
restriction halves the state dimension and is the reason full rank is \num{4}
rather than \num{6}; in three dimensions the rank deficiency argument is
unchanged in kind, since range still has no velocity dependence, but the
arc-length threshold and the conditioning magnitudes would differ. The circular
restriction omits lunar orbital eccentricity, solar perturbation and
non-spherical gravity, each of which perturbs the flow that carries the
velocity information, so the threshold reported here should be read as
characteristic of the regime rather than as a prediction for a specific
mission.

Measurements are noise-free geometric ranges. There is no light-time
correction, no clock error, no measurement bias and no outage modelling. This
is the right choice for an observability study, because observability is a
property of the deterministic measurement-and-dynamics pair and adding noise
would obscure a rank question with a covariance question, but it means the
condition numbers are lower bounds on the difficulty of a real estimation
problem rather than predictions of achieved accuracy.

The reference spacecraft are treated as having known states. In a genuine
linked autonomous navigation problem all members are unknown simultaneously,
which enlarges the state and changes both the rank bookkeeping and the
conditioning. We chose the simpler formulation deliberately, because it isolates
the mechanism the paper is about, but the joint problem is the operationally
relevant one and is the natural next step.

Finally, the constellation is one four-member family at one set of phases. We
have not swept over constellation size, orbit family, or relative phasing, so we
report no claim about optimal geometry. The claims we do make, the structural
rank deficiency, the existence of an arc threshold, the monotone improvement of
conditioning with arc, and the Doppler lever, are consequences of the
measurement model and the dynamics and do not depend on that particular choice.

\section{Conclusion}
\label{sec:conclusion}

Dilution of precision is the wrong instrument for cislunar navigation, not
because the geometry is poor but because the quantity is undefined: range
measurements carry no velocity information, so the instantaneous Jacobian of a
range-tracked spacecraft is structurally rank-deficient against its own state,
at rank \num{2} of \num{4} for the constellation studied here. Observability is
supplied instead by the three-body flow acting over time.

Measured that way, the problem has a clean threshold. The full state becomes
observable after a tracking arc of \SI{2.09}{\hour}, a figure at which a square-root information filter, recomputing the same quantity by a different route, becomes well-posed at the same arc
with condition numbers agreeing to six significant figures. Past that
threshold, conditioning rather than rank is the design axis, improving by about
six orders of magnitude between \SI{3}{\hour} and \SI{48}{\hour} of arc.

The most actionable result is the measurement-type lever. Adding range rate to
the same links lifts the instantaneous rank to full, so a Doppler-capable
inter-satellite link converts a problem requiring hours of continuous tracking
into one with a snapshot solution. For an architecture being specified now,
that is a cheaper and more decisive choice than adding spacecraft, and it is
invisible to any analysis conducted in terms of dilution of precision.

\section*{Data and code availability}
\label{sec:availability}
\addcontentsline{toc}{section}{Data and code availability}

All results are produced by the open Kshana PNT engine
\citep{baweja2026kshana} through its \code{cislunar-observability} scenario
kind, which supplies the three-body propagation, the variational state
transition matrix, the differential corrector, the observability Gramian and
the square-root information filter used as a second computational route. The
scenario definitions, the arc-length sweep, the per-orbit periodicity residuals
of Table~\ref{tab:dro} and the conditioning table behind
Table~\ref{tab:arc} and Figure~\ref{fig:main} are released as machine-readable
files alongside the manuscript. The sweeps are deterministic and involve no
random draws.

\section*{Acknowledgements}
\addcontentsline{toc}{section}{Acknowledgements}
This work is part of the Kshana program on reproducible, honestly-scoped PNT
analysis.

\bibliographystyle{plainnat}
\bibliography{cislunar-observability}
\end{document}